\documentclass{article}
\usepackage{graphics} 
\begin{document}
\title{First experimental tests of the Peyrard-Bishop model
   applied to the melting of very short DNAs}
\author{Alessandro Campa\thanks{Author to whom correspondence should
         be addressed. E-mail:campa@axiss.iss.infn.it} \\
  \small{$^1$Physics Laboratory, Istituto Superiore di Sanit\`a
         and INFN Sanit\`a} \\
  \small{Viale Regina Elena, 299, 00161 Roma, Italy}
 \and
        Andrea Giansanti \\
  \small{$^2$Physics Dept., Universit\`a di Roma ``La Sapienza''
         and INFM Unit\`a Roma 1} \\
  \small{P.le Aldo Moro, 2, 00185 Roma, Italy}}
\date{February 23, 1998}
\maketitle

\begin{abstract}
The melting curves of short heterogeneous DNA chains in solution are 
calculated on the basis of statistical thermodynamics and compared
to experiments.  The computation of the partition function is
based on the Peyrard-Bishop  hamiltonian, which has already been adopted in
the theoretical description  of the melting of long DNA chains. In the case
of short chains it is necessary to consider not only the breaking of the
hydrogen bonds between single base pairs, but also the
complete dissociation of the two strands forming the double helix.
\end{abstract}

There is a need for a theory of the melting of short DNA chains
(oligonucleotides).
The melting is the highly cooperative thermal disruption
of the hydrogen bonds between complementary bases in the double helix,
 as usually
monitored by the UV absorption increment due to the unstacking
of the separated bases \cite{cant}. At
the equilibrium melting temperature half of the bonds are disrupted.
Synthetic oligonucleotides of a fixed length and base pairs sequence
have been used for a long time as model systems for the study of
the structural and thermodynamical properties of the longer and 
more complex natural forms of DNA \cite{marky83}. Many studies have shown the
effects of both sequence and solvent composition on the melting curves   
of oligonucleotides in solution \cite{Bre}.
More recently particular attention has been given to the study
of sequence specific effects on the thermal stability of a variety of
specially designed oligonucleotides, due to their importance in the
exploitation
of molecular biological techniques in gene therapy \cite{kool96} and genome
mapping \cite{doktycz95}. Predictive information has been gained through
an extensive thermodynamical investigation on the melting behavior of
oligonucleotides, based on the computation of the Gibbs free energy,
at a fixed solvent composition, as a sum of contributions from
nearest neighbors in the sequences \cite{Bre86,sant}.
This phenomenology and the predictive power of the thermodynamical
approach should then be
confronted with a microscopic theory of short, heterogeneous DNA chains.

Modellization of DNA melting was initially motivated by the study of
the important process of transcription, in which the double
helix has to be locally opened to allow reading of the genetic code.
It was based, already many years ago, on Ising-like models \cite{pol,wart}, 
and more recently on an approach based on
the modified self-consistent phonon approximation \cite{dp95} (see also
\cite{pro} and references therein). These methods allow only equilibrium
estimates of the probability of bond disruption. However, it is also
important to consider DNA dynamics, both at melting and pre-melting
temperatures. There is an interest in relaxation and kinetic phenomena,
which are relevant for the pharmacological applications \cite{kool96},
and in the study of nonlinear energy localization and transduction. With a
particular focus on the last problem, 
discrete nonlinear models of DNA (see, e. g., \cite{yak1,yak2}, and for a
review \cite{gae}), have been introduced;
sequence effects have been considered in \cite{sal}. These models are
appealing, because they are simplified microscopic models with a small
number of degrees of freedom, and thus are affordable also for the
simulation of very long times. The experimentally available melting curves
offer a way to optimize the parameters of these models, and therefore also
increase the confidence for their use in dynamical studies.

With a particular interest in thermal
stability, a dynamical model was introduced by Peyrard and Bishop in
1989 \cite{pb89} (PB model).
The authors have shown, through statistical mechanics calculations
and constant
temperature molecular dynamics \cite{pb89,dpb1,dpb2}, applied to the case
of a very long homogeneous DNA chain, that the model can give
a satisfactory melting curve, especially after the improvement introduced
in \cite{dpb2}. The PB model has been successively applied to heterogeneous
chains, either modelling the heterogeneity with a quenched
disorder \cite{cul}, or properly choosing basis sets of orthonormal
functions for the kernels appearing in the expression of the partition
function \cite{zhan}, but comparison with experimental data was not
attempted. In all these works the fact that the DNAs considered are quite
long was essential, for the following reason.
In a solution with two types of DNA single strands, $A$ and $B$, there is
a thermal equilibrium between dissociated strands and associated double
strands (the duplexes $AB$), and a thermal equilibrium, in the duplexes,
between broken and unbroken interbase hydrogen bonds. 
The average fraction $\theta$ of bonded base pairs can then be factorized
as $\theta = \theta_{ext}\theta_{int}$ \cite{pol,wart}. $\theta_{ext}$
is the average fraction of strands forming duplexes, while
$\theta_{int}$ is the average fraction of unbroken bonds in the duplexes.
The dissociation equilibrium can be neglected in the case of long
chains, where $\theta_{int}$ and thus $\theta$ go to
0 when $\theta_{ext}$ is still practically 1. On the contrary,
in the case of short chains the
processes of single bond disruption and strand dissociation tend to happen
in the same temperature range; therefore, the computation of both
$\theta_{int}$ and $\theta_{ext}$ is essential. In Ref. \cite{zhan} the
factorization of $\theta$ is stated, but only the case of long chains is
then considered.

The aim of this work is to show, through a comparison with experimental data,
that the onedimensional PB model can be used to compute the melting curves of
short DNAs. It will also be shown how to take into account the
dissociation equilibrium.

The potential of the PB model \cite{pb89,dpb1,dpb2}
is given by:
\begin{equation}
U=\sum_i \Bigl\{\frac{k}{2}\left[1+\rho e^{-\alpha(y_{i+1}+y_i)}\right]
{(y_{i+1}-y_i)}^2 +D_i{\left(e^{-a_i y_i}-1\right)}^2\Bigr\},
\label{hamil}
\end{equation}
where $y_i$ is the distance between the $i$-th complementary bases
minus their equilibrium separation.
The parameters
$k$, $\rho$ and $\alpha$ refer to the anharmonic stacking
interaction, while the interbase bond is represented
by a Morse potential, with depth $D_i$ and width $a_i$.
In Refs. \cite{pb89,dpb1,dpb2} there is only a single
parameter $D$ because only homogeneous DNAs have been considered.
The stacking interaction, that in
the first attempts \cite{pb89,dpb1} was purely harmonic ($\rho=0$), decreases
when the complementary bases get farther ($\rho$ positive): this $\rho$
dependent nonlinear term was found to be relevant to give cooperativity
to the melting process \cite{dpb2}.

To model heterogeneous DNAs, we have inserted two different values
of $D_i$, according to the two possible Watson-Crick base pairs:
adenine-thymine (A-T) and guanine-cytosine (G-C). The former has
two hydrogen bonds, while the latter has three. We have then
chosen a depth for
the G-C Morse potential 1.5 times that for the A-T Morse potential. The
complete set of parameter values that we have chosen is : $k=0.025$
eV/\AA$^2$, $\rho=2$, $\alpha=0.35$ \AA$^{-1}$, $D_{AT}=0.05$ eV,
$D_{GC}=0.075$ eV, $a_{AT}=4.2$ \AA$^{-1}$, $a_{GC}=6.9$ \AA$^{-1}$. These
values have been adjusted to
reproduce the experimentally observed melting temperature of long
homogeneous DNA in the most usual solvent conditions \cite{wart,note1}.
For a given set of values, the melting temperatures can be deduced
with the technique of the transfer matrix method \cite{pb89,dpb1,dpb2}.

We have then made a
statistical mechanics computation, in which partition functions have
been used to obtain both $\theta_{int}$ and $\theta_{ext}$.
For the computation of $\theta_{int}$ one has to separate the
configurations describing a double strand on the one hand, and dissociated
single strands on the other. The very possibility of dissociation
makes this a non trivial problem. We have adopted the following
strategy. The $i$-th bond is considered disrupted if the value of
$y_i$ is larger than a chosen threshold $y_0$. We have therefore
defined a configuration to belong to the double strand if at least
one of the $y_i$s is smaller than $y_0$. It is then natural to define
$\theta_{int}$ for an $N$ base pair duplex by:
\begin{displaymath}
\theta_{int}=\frac{1}{N}\sum_{i=1}^N<\vartheta(y_0-y_i)>
\end{displaymath}
where $\vartheta(y)$ is the Heaviside step function and the canonical
average $<\cdot>$ is defined considering only the double strand
configurations. We have chosen a value of 2 \AA \ for $y_0$.
After a discretization of the coordinate
variables and the introduction of a proper cutoff on the maximum
value of the $y_i$s \cite{dp95}, the computations
needed for the canonical averages are readily reduced to the multiplication
of finite matrices, since the potential (\ref{hamil}) couples only nearest
neighbors, and are easily performed by suitable computer programs.

Let us now consider $\theta_{ext}$.
At equilibrium the chemical potentials
of the three species $A$, $B$ and $AB$ \cite{note2}
are related by the equation:
$\mu_{AB}-\mu_A-\mu_B = 0$.
Using the definition of the chemical potentials as
derivatives of the free energy, and in turn the relation
of the latter to the partition functions,
we obtain an equation involving appropriate partition
functions. In the usual experimental conditions the solutions
can be considered ideal; with the further assumption that the model takes
into account effectively the presence of the solvent, we get
the usual equilibrium condition:
\begin{displaymath}
\frac{N_{AB}Z(A)Z(B)}{N_AN_BZ(AB)}=1
\end{displaymath}
where $N_j$ is the number of molecules of species $j$ in the volume $V$
considered, and $Z(j)$ is the partition function of a molecule of species
$j$ in V \cite{note3}. The numbers $N_j$ are related by the constraints
$2N_{AB}+N_A+N_B=const\equiv 2N_0$ and
$\Delta N_A=\Delta N_B=-\Delta N_{AB}$. Considering the case $N_A=N_B$
(the experimental curves that we are presenting are made in these
conditions, with the duplex obtained by annealing equal concentrations
of $A$ and $B$), we arrive at the following expression for
$\theta_{ext}\equiv N_{AB}/N_0$:
\begin{displaymath}
\theta_{ext}=1+\delta-\sqrt{\delta^2+2\delta}
\end{displaymath}
where $\delta$ is given by the following expression:
\begin{equation}
\delta=\frac{Z(A)Z(B)}{2N_0Z(AB)}\equiv\frac{Z_{int}(A)Z_{int}(B)}
{a_{av}Z_{int}(AB)}\frac{a_{av}Z_{ext}(A)Z_{ext}(B)}{2N_0Z_{ext}(AB)},
\label{separ}
\end{equation}
where in the rightmost side we have introduced
the separation of the partition functions in an internal and an external
part \cite{pol,wart}; the meaning of $a_{av}$ will be explained in a moment.
For the calculation of the internal functions,
that do not include the overall translation of the molecules,
we use the DNA model described above (which is also simply adapted
to the description of single strands, allowing an analytical evaluation:
only a harmonic stacking
interaction remains, which is weaker than in the duplex, since in this
case the term involving $\rho$ is 0). We have chosen to insert in the last
side of Eq.\ (\ref{separ}) $a_{av}=\sqrt{a_{AT}a_{GC}}$ to make separately
dimensionless both fractions, that therefore can not depend on the choice of
units. Without any such normalization the first fraction would have the
dimensions of an inverse of a length, since the overall translation is not
included in $Z_{int}$. It is included in the external functions, that,
however, have to take into account also the
dynamics not described by the simple onedimensional model, and related
to conformational movements (like, for example, the
winding of the strands). This point has already been considered in
Ising models: the influence on the dissociation process of the degrees of
freedom not described by the model can not be neglected,
and it must be accounted for in some way. In analogy to what has been
proposed for the Ising models \cite{pol,wart} on the basis of the
partition functions of rigid bodies \cite{land}, we make the following
choice:
\begin{equation}
\frac{a_{av}Z_{ext}(A)Z_{ext}(B)}{2N_0Z_{ext}(AB)}
=\frac{n^*}{n_0}N^{-p\theta_{int}+q}
\label{extpar}
\end{equation}
where the parameters $p$ and $q$ can be fixed by a comparison with
experimental melting curves; $n_0$ is the single strand
concentration $N_0/V$, and $n^*$ is a chosen reference concentration
(we have taken 1 $\mu$M, a usual concentration in experiments).
We defer further comments about this equation
after the presentation of the results.

\begin{figure}[h]
\includegraphics[40mm,212mm][180mm,295mm]{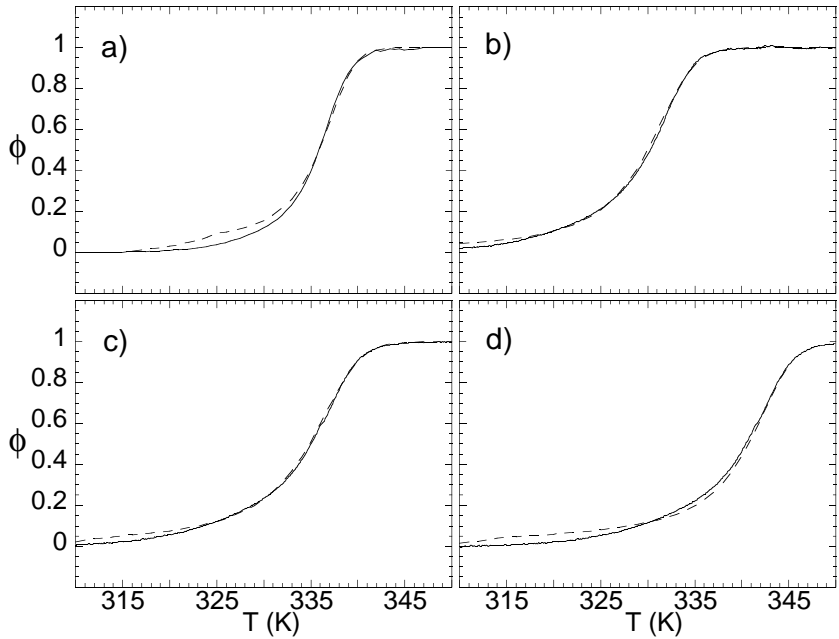}
\caption{Experimental melting profiles (full lines) and theoretical
results (dashed lines) for the three DNA chains. We have plotted the value
of $\phi\equiv 1-\theta$. Panel a): sequence $s_1$; panel b): sequence $s_2$;
panel c): sequence $s_3$ at the lower concentration; panel d): sequence $s_3$
at the higher concentration. The fitted parameters $p$ and $q$ have the
following values: $p=32.43$ and $q=29.30$ for $s_1$; $p=36.77$ and $q=34.89$
for $s_2$; $p=29.49$ and $q=27.69$ for $s_3$.}
\label{fig1} 
\end{figure}

We show here the comparison of our calculations with the experimental
melting curves that have been obtained, in our lab, for three different
oligonucleotides, in a 10mM Na phosphate buffer, 0.1 mM Na$_2$EDTA,
200 mM NaCl, pH 6.7.
One of the oligonucleotides contained 27 base pairs, and the other
two had 21 base pairs. The sequences are given by:
\begin{eqnarray*}
s_1)&&\,\,{\rm ^{5'}CTTCTTATTCTTATTGTTCGTCTTCTC_{3'}} \\
s_2)&&\,\,{\rm ^{5'}CTCTTCTCTTCTTTCTCTCTC_{3'}} \\
s_3)&&\,\,{\rm ^{5'}GTGTTAACGTGAGTATAGCGT_{3'}}
\end{eqnarray*}
and by the respective complementary strands. We have considered the case
$s_3)$ at two different concentrations. The single strand concentration was:
$s_1)$: 2.4 $\mu$M, $s_2)$: 1.7 $\mu$M, $s_3)$: 3.1 $\mu$M and 120 $\mu$M.
In Fig.\ \ref{fig1} we show the experimental and computed melting curves.
As it can be seen, there are sequence and concentration effects on the
experimental melting curves, which are well reproduced by the computed
curves. Note that a 40 fold concentration increase for $s_3)$ yields an
increase of only 5 degrees in the melting temperature (a logarithmic
dependence on the concentration is expected \cite{cant}). Similar
differences between curves at the low concentrations should then be due
to sequence and length effects.
We would like to stress that
in the case $s_3)$ the parameters $p$ and $q$ have been fitted to the
experimental curve at the lower concentration. The comparison with the
experimental curve at the higher concentration has then been performed
with only the change of the value of $n_0$ in Eq.\ (\ref{extpar}),
without changing the values of $p$ and $q$; this has reproduced the
difference between the melting temperatures of the two cases, that
differ by about 5 degrees. This fact indicates that the concentration
dependence of the left hand side of Eq.\ (\ref{extpar}) is described
by the preexponential factor, while the parameters $p$ and $q$
are related to the molecular conformation.

In conclusion, our comparisons show that it is feasible to compute
the equilibrium melting profile of DNA oligonucleotides with
the PB nonlinear model. We would also like to note that
the modellization of the external partition functions ratio as in
Eq.\ (\ref{extpar}) is very similar to that adopted in Ising models for
medium size DNAs (100-600 base pairs) \cite{pol,wart}. This confirms that
this term is related to the conformational flexibility of the
double and single strands, not described by a onedimensional model.
The internal term is related to the onedimensional hamiltonian and
then to nearest neighbor interactions. For long
DNAs (large $N$), at temperatures in which $\theta_{int}$ is already close
to 0, the part in Eq.\ (\ref{separ}) depending on the internal partition
functions goes as $e^{-\gamma N}$ for some positive $\gamma$, and thus
$\delta \approx 0$ and $\theta_{ext}\approx 1$. This $N$ dependence of the
internal part can be seen, for example, in the case of homogeneous sequences
with the transfer matrix method \cite{pb89,dpb1,dpb2}. It is expected to be
the same for heterogeneous sequences.

In very
short chains like ours, it is not surprising that the specific sequence has
some influence on the parameters $p$ and $q$, while in medium chains some
self-averaging effects should already take place. In fact, as shown in the
caption to Fig.\ \ref{fig1}, we have found differences of about 25 percent
in the parameters referring to different sequences. We are now working on a
more extended set of melting curves for a properly chosen set of
oligonucleotides, that can help in the attempt to find the relation
between the specific sequence and the optimized parameters. Then it would
be possible to test the predictive power of this model and confront it
with the predictions of purely thermodynamical calculations.

In a more extended paper in preparation we will show a more exhaustive
comparison with experimental curves. We will also check if a simple
analysis based on the number of occurrences of the different intrastrand
nearest neighbor couples in the sequences is sufficient to obtain the
parameters,
similarly to what happens in the calculations of Gibbs free energy in
short oligonucleotides \cite{Bre86,sant}.

We are very grateful to F. Barone, M. Matzeu, F. Mazzei and F. Pedone
for providing the experimental melting curves and for illuminating
discussions.


\begin{thebibliography}{99}
\bibitem{note1}
These temperatures differ by about 40 $K$, and their
exact values depend on the solvent conditions, especially
the ionic strength of the solution.
This means that in any effective model, dynamical or thermodynamical,
the parameters should depend on the solvent conditions.
\bibitem{note2}
We are considering non self-complementary strands, i. e., $A\ne B$.
\bibitem{note3}
To reproduce the experimental conditions one should use
isobaric-isothermal (P,T), rather than canonical isochoric-isothermal
(V,T), partition functions. The practical ideality of the solution allows
the use of the latter without appreciable differences.
\end{thebibliography}
\end{document}